\documentclass[11pt,usletter]{article}

\usepackage{amsthm}
\usepackage{amsmath}
\usepackage{breqn}

\usepackage{jheppub}
\pdfoutput=1

\usepackage{graphicx}
\usepackage{hyperref}
\usepackage{color}
\usepackage{epsfig}
\usepackage{amsfonts}
\usepackage{amssymb}
\usepackage{amsmath}
\usepackage{amsthm}
\usepackage{subfig}
\usepackage{bm}
\usepackage{mathrsfs}
\usepackage{bbm}
\usepackage{cancel}
\usepackage{accents}

\def\ba{\begin{align}}
\def\ea{\end{align}}
\def\be{\begin{equation}}
\def\ee{\end{equation}}
\def\bea{\begin{eqnarray}}
\def\eea{\end{eqnarray}}
\usepackage[utf8]{inputenc}
\usepackage[english]{babel}

\begin{document}

\title{Quantum complexity and the virial theorem}
\subheader{\rm \hfill CALT-TH-2018-016}
\date{\today}

\author[a]{Ning Bao,}
\author[b,c]{Junyu Liu}

\affiliation[a]{Berkeley Center for Theoretical Physics,\\ University of California, Berkeley, CA 94720, USA}
\affiliation[b]{Walter Burke Institute for Theoretical Physics,\\ California Institute of Technology, Pasadena, California 91125, USA}
\affiliation[c]{Institute for Quantum Information and Matter,\\ California Institute of Technology, Pasadena, California 91125, USA}

\emailAdd{ningbao75@gmail.com}
\emailAdd{jliu2@caltech.edu}

\abstract{It is conjectured that in the geometric formulation of quantum computing, one can study quantum complexity through classical entropy of statistical ensembles established non-relativistically in the group manifold of unitary operators. The kinetic and positional decompositions of statistical entropy are conjectured to correspond to the Kolmogorov complexity and computational complexity, respectively, of corresponding quantum circuits. In this paper, we claim that by applying the virial theorem to the group manifold, one can derive a generic relation between Kolmogorov complexity and computational complexity in the thermal equilibrium.
}


\maketitle
\flushbottom
\section{Introduction}
Quantum complexity, a concept which determines time or quantum gate cost to achieve given problems in computer science, has addressed several attentions in the high energy theory community. Based on holographic duality, a theoretical correspondence between bulk gravity and boundary field theories, the boundary complexity is claimed to be equal to gravity action   in some circumstances, and the growth of complexity in the boundary could be evaluated by computations of gravity action \cite{Brown:2015bva,Brown:2015lvg,Brown:2016wib}. (See also, some related works in this area \cite{Chapman:2016hwi,Carmi:2016wjl,Jefferson:2017sdb,Carmi:2017jqz,Bernamonti:2018vmw,Fu:2018kcp,Chapman:2018dem,Chapman:2018lsv,Hackl:2018ptj})
\\
\\
One motivation to think boundary complexity is a possible equivalent object with the gravity action is based on the following geometric observation from Nielsen \cite{Ne0,Ne}. Given an $K$-qubit system, one could consider evolution operators as points living in a group manifold $\text{SU}(2^K)$. One can further assign a metric to this manifold, whereupon under a certain manipulation the complexity of the corresponding operator could be understood as the geodesic length that connects the operator and the identity. This idea provides a novel geometric way to study complexity theory, and has led to deep conjectures relating the concepts of quantum complexity, holographic duality, and the nature of quantum gravity.
\\
\\
The geometric version of quantum complexity naturally connects quantum physics of $k$-local Hamiltonians with the notion of the disordered average and statistical motion of particle moving in a group manifold. In this sense, one can conjecture a relation between statistical entropy and quantum complexity. Following the second law of thermodynamics, a similar second law holding for complexity growth has also been conjectured \cite{Brown:2017jil}. More precisely, the conjecture states that one can decompose the entire statistical entropy of classical particle in a group manifold into kinetic and positional parts, where the former corresponds to \emph{Kolmogorov complexity}, and the latter the \emph{computational complexity} of the corresponding quantum system.
\\
\\
The Kolmogorov complexity is roughly speaking the minimal cost to specify bit strings, while the computational complexity is the cost of time or the scale of depth for quantum circuits. As two different complexity measures, it is natural to ask if there exists some possible connections between them. The entropic  conjectures about complexity provide us a different angle on this problem, where in the dual classical system, the physics could be understood more intuitively by addressing the property of statistical entropies.
\\
\\
Working in the canonical ensemble, the kinetic-positional decomposition of the statistical entropy is naively the decomposition of the whole Hamiltonian into kinetic and potential energy. In this case, from basic properties of mechanics, or more fundamentally, the equation of motion, one could naturally expect there to potentially be a relation between statistical average of potential and kinetic energy. In ordinary classical and statistical mechanics, the direct answer is celebrated virial theorem.
\\
\\
In this paper, we will study this problem by analyzing a modified version of the virial theorem on the group manifold $\text{SU}(2^K)$. By working directly in the curved geometry, one can arrive at a modified version of the usual virial theorem, where the average of potential and kinetic energies are related by the affine connection terms of the curved space. Thus, connecting with the arguments identifying complexity with entropy, we show a natural relation between two notions of complexities in quantum information theory.
\\
\\
This paper is organized as follows. In Section \ref{virial}, we discuss the extension of the virial theorem to curved space. In Section \ref{relation}, we discuss the relationship between entropies, and alternatively, complexities as a consequence of this modified virial theorem. In Section \ref{con}, we conclude and discuss some possible future directions related to this research.
\section{Virial theorem}\label{virial}
\subsection{Traditional virial theorem}
The (classical) virial theorem is a connection between the potential and kinetic energy of a statistical system. Here we will review the derivation in classical mechanics as a warmup.
\\
\\
For a particle system with location $\mathbf{r}_i$ and mass $m_i$ we have momenta
\begin{align}
{{\mathbf{p}}_{i}}=m_i\frac{\partial {{\mathbf{r}}_{i}}}{\partial t}
\end{align}
and thus can define the function
\begin{align}
G=\sum\limits_{i}{{{\mathbf{p}}_{i}}\cdot {{\mathbf{r}}_{i}}}
\end{align}
Further, we have that
\begin{align}
  & \frac{\partial G}{\partial t}=\sum\limits_{i}{\frac{\partial {{\mathbf{p}}_{i}}}{\partial t}\cdot {{\mathbf{r}}_{i}}}+\sum\limits_{i}{{{\mathbf{p}}_{i}}\cdot \frac{\partial {{\mathbf{r}}_{i}}}{\partial t}} \nonumber\\
 & =\sum\limits_{i}{\frac{\partial }{\partial t}\left( {{m}_{i}}\frac{\partial {{\mathbf{r}}_{i}}}{\partial t} \right)\cdot {{\mathbf{r}}_{i}}}+\sum\limits_{i}{{{m}_{i}}\frac{\partial {{\mathbf{r}}_{i}}}{\partial t}\cdot \frac{\partial {{\mathbf{r}}_{i}}}{\partial t}} \nonumber\\
 & =\sum\limits_{i}{{{m}_{i}}\left( \frac{{{\partial }^{2}}{{\mathbf{r}}_{i}}}{\partial {{t}^{2}}} \right)\cdot {{\mathbf{r}}_{i}}}+2E_K
\end{align}
Here we define
\begin{align}
E_K=\sum\limits_{i}{\frac{1}{2}{{m}_{i}}\frac{\partial {{\mathbf{r}}_{i}}}{\partial t}\cdot \frac{\partial {{\mathbf{r}}_{i}}}{\partial t}}
\end{align}
to be the kinetic energy. Now the force is given by Newton's law
\begin{align}
{{m}_{i}}\left( \frac{{{\partial }^{2}}{{\mathbf{r}}_{i}}}{\partial {{t}^{2}}} \right)={{\mathbf{F}}_{i}}
\end{align}
So we have
\begin{align}
\frac{\partial G}{\partial t}=\sum\limits_{i}{{{\mathbf{F}}_{i}}\cdot {{\mathbf{r}}_{i}}}+2E_K
\end{align}
If the system is a stably bound system, we have the derivative of $G$ vanishes after time average\footnote{For a stable, bounded system, after the time average we have
\begin{align}
\left\langle \frac{\partial G}{\partial t} \right\rangle \sim \underset{t\to \infty }{\mathop{\lim }}\,\frac{G(t)-G(0)}{t}\le \underset{t\to \infty }{\mathop{\lim }}\,\left| \frac{\max (G)-\min (G)}{t} \right|=0
\end{align}
}, thus giving us
\begin{align}
\left\langle \sum\limits_{i}{{{\mathbf{F}}_{i}}\cdot {{\mathbf{r}}_{i}}} \right\rangle =-2\left\langle E_K \right\rangle
\end{align}
This naturally relates force to potential energy. We know that typically the potential is a function depending only on distance of particles, namely, that the Lagrangian is
\begin{align}
L=E_K-V=\left( \sum\limits_{i}{\frac{1}{2}{{m}_{i}}\frac{\partial {{\mathbf{r}}_{i}}}{\partial t}\cdot \frac{\partial {{\mathbf{r}}_{i}}}{\partial t}} \right)-V
\end{align}
where the potential $V$ should only depend on positions. So the force is given by
\begin{align}
{{\mathbf{F}}_{i}}=-\frac{\partial V}{\partial {{\mathbf{r}}_{i}}}
\end{align}
resulting in the virial theorem:
\begin{align}
\left\langle \sum\limits_{i}{\frac{\partial V}{\partial {{\mathbf{r}}_{i}}}\cdot {{\mathbf{r}}_{i}}} \right\rangle =2\left\langle E_K \right\rangle
\end{align}
While this version of virial theorem works for time average of particle trajectories, one can derive a similar result from statistical ensembles, such as, for instance, the canonical ensemble.
\\
\\
Now let us consider a system with $N$ particles moving in $d$-dimensional flat space. The dimension of the phase space in this case is $2dN$. We can write the indices collectively as $a,b,\text{etc.}=1,2,\cdots,dN$. Now consider the quantity
\begin{align}
\left\langle {{x}_{a}}\frac{\partial H}{\partial {{x}_{b}}} \right\rangle =C\int{dxdp{{e}^{-\beta H}}{{x}_{a}}\frac{\partial H}{\partial {{x}_{b}}}}
\end{align}
where here $C$ is the normalization constant of the expectation value. One finds that
\begin{align}
  & C\int{dxdp\left( {{e}^{-\beta H}}{{x}_{a}}\frac{\partial H}{\partial {{x}_{b}}} \right)}=-\frac{C}{\beta }\int{dxdp\left( {{x}_{a}}\frac{\partial {{e}^{-\beta H}}}{\partial {{x}_{b}}} \right)} \nonumber\\
 & =\frac{C}{\beta }\int{dxdp\left( \frac{\partial {{x}_{a}}}{\partial {{x}_{b}}}{{e}^{-\beta H}} \right)}=\frac{C}{\beta }{{\delta }_{ab}}\int{dxdp\left( {{e}^{-\beta H}} \right)}=\frac{1}{\beta }{{\delta }_{ab}}
\end{align}
specifically, we know that, taking $a=b$, we have that
\begin{align}
\left\langle {{x}_{a}}\frac{\partial H}{\partial {{x}_{a}}} \right\rangle =\frac{1}{\beta }
\end{align}
Note that the above expression has no sum. The same logic applies if one replaces $x$ by $p$, obtaining
\begin{align}
\left\langle {{p}_{a}}\frac{\partial H}{\partial {{p}_{a}}} \right\rangle =\frac{1}{\beta }
\end{align}
As a conclusion we get
\begin{align}
\left\langle {{x}_{a}}\frac{\partial H}{\partial {{x}_{a}}} \right\rangle =\left\langle {{p}_{a}}\frac{\partial H}{\partial {{p}_{a}}} \right\rangle
\end{align}
This is the statistical version of the virial theorem, which is more constraining than the mechanical one (because the statistical version fixes the ensemble). We can see this by applying the Hamilton equation
\begin{align}
  & \sum\limits_{a}{\left\langle {{x}_{a}}\frac{\partial H}{\partial {{x}_{a}}} \right\rangle }=\sum\limits_{a}{\left\langle {{x}_{a}}\frac{\partial V}{\partial {{x}_{a}}} \right\rangle } \nonumber\\
 & \sum\limits_{a}{\left\langle {{p}_{a}}\frac{\partial H}{\partial {{p}_{a}}} \right\rangle }=\sum\limits_{a}{\left\langle {{p}_{a}}{{{\dot{x}}}_{a}} \right\rangle }=\left\langle 2E_K \right\rangle
\end{align}
So we could see that, in the case of flat space, these two versions lead to the same result. The consistency between the time average and the ensemble average is a consequence of ergodicity.
\subsection{Virial theorem in the curved space}
\subsubsection{Setup}
Now we will discuss the virial theorem in the curved space. Existing literature has extended the virial to the curved spacetime in the language of relativity with the application of astrophysics, for instance, in the context of studying dark matter (see \cite{Chandrasekhar:1965gcg,B1,K,B2,B3}). However, currently we are interested in only the non-relativistic case, where the goal is to study trajectories of particles moving in a general curved space instead of spacetime, as this is the problem relevant for studying the Nielsen complexity geometry.
\\
\\
We start by considering the Lagrangian in the curved space. Let the space $\mathcal{M}$ be a Euclidean manifold. The coordinate of particle $i$ is denoted by $x_i^\mu$, where $\mu$ are the indices for vectors on the manifold. The metric on $\mathcal{M}$ is given by $g_{\mu\nu}$. We know that the Lagrangian for many free particles labeled by $i$ is given by its kinetic energy
\begin{align}
E_K=\frac{1}{2}\sum\limits_{i}{{{m}_{i}}{{g}_{\mu \nu }}({{x}_{i}})\dot{x}_{i}^{\mu }}\dot{x}_{i}^{\nu }
\end{align}
The whole Hamiltonian is
\begin{align}
  & H=\sum\limits_{i}{\frac{1}{2}{{m}_{i}}{{g}_{\mu \nu }}({{x}_{i}})\dot{x}_{i}^{\mu }\dot{x}_{i}^{\nu }}+V({{x}_{i}}) \nonumber\\
 & =\sum\limits_{i}{\frac{1}{2{{m}_{i}}}{{g}_{\mu \nu }}({{x}_{i}})p_{i}^{\mu }p_{i}^{\nu }}+V({{x}_{i}})
\end{align}
where the momenta are defined by
\begin{align}
p_{i}^{\mu }={{m}_{i}}\dot{x}_{i}^{\mu }
\end{align}
Now we can also define curved phase space. The positional part of the phase space is given by the manifold coordinates, while the element volume is given by the invariant volume
\begin{align}
d{{V}_{x,i}}=\sqrt{g({{x}_{i}})}\prod\limits_{\mu }{dx_{i}^{\mu }}
\end{align}
For given $x$, $p$ is located in the tangent space of $x$. Therefore, for any $x$, $p$ lives in flat space. If we set $\text{dim}(\mathcal{M})=d$, then the space for momentum to be integrated over is $\mathbb{R}^d$. So we define the phase space to be
\begin{align}
{{\left( \mathcal{M}\times {{\mathbb{R}}^{d}} \right)}^{N}}
\end{align}
with volume element
\begin{align}
d\Omega=\prod\limits_{i}{d{{\Omega}_{x,i}}d{{\Omega}_{p,i}}}=\prod\limits_{i}{\sqrt{g({{x}_{i}})}\prod\limits_{\mu ,\mu '}{dx_{i}^{\mu }dp_{i}^{\mu '}}}
\end{align}
For a canonical ensemble we have the inverse temperature $\beta$, giving a phase factor of
\begin{align}
P(x,p)=\exp (-\beta H)=\exp \left( -\beta \left( \sum\limits_{i}{\left( \frac{1}{2{{m}_{i}}}{{g}_{\mu \nu }}({{x}_{i}})p_{i}^{\mu }p_{i}^{\nu } \right)}+V({{x}_{i}}) \right) \right)
\end{align}
In particular, one can decompose it as
\begin{align}
P(x,p)=\exp (-\beta H)=\exp (-\beta V)\exp (-\beta E_K)
\end{align}
\subsubsection{Statistical version}
Now we start to derive a virial theorem for a canonical ensemble. We begin by considering how to interpret the expression
\begin{align}
\left\langle x_{i}^{\mu }\frac{\partial H}{\partial x_{j}^{\nu }} \right\rangle =C\int{\exp (-\beta H)x_{i}^{\mu }\frac{\partial H}{\partial x_{j}^{\nu }}d\Omega }
\end{align}
One could write it as
\begin{align}
  & \left\langle x_{i}^{\mu }\frac{\partial H}{\partial x_{j}^{\nu }} \right\rangle =-\frac{C}{\beta }\int{x_{i}^{\mu }\frac{\partial {{e}^{-\beta H}}}{\partial x_{j}^{\nu }}d\Omega } \nonumber\\
 & =\frac{1}{\beta }\delta _{\nu }^{\mu }{{\delta }_{ij}}+\frac{C}{\beta }\int{x_{i}^{\mu }{{e}^{-\beta H}}\frac{\partial d\Omega }{\partial x_{j}^{\nu }}}
\end{align}
Using the formula
\begin{align}
{{\partial }_{\mu }}g=g{{g}^{\alpha \beta }}{{\partial }_{\mu }}{{g}_{\alpha \beta }}
\end{align}
we get
\begin{align}
\frac{\partial d\Omega }{\partial x_{j}^{\nu }}=\frac{d\Omega }{2}{{g}^{\alpha \beta }}({{x}_{j}}){{\partial }_{\nu }}{{g}_{\alpha \beta }}({{x}_{j}})
\end{align}
So we get
\begin{align}
\left\langle x_{i}^{\mu }\frac{\partial H}{\partial x_{j}^{\nu }} \right\rangle =\frac{1}{\beta }\delta _{\nu }^{\mu }{{\delta }_{ij}}+\frac{1}{2\beta }\left\langle x_{i}^{\mu }{{g}^{\alpha \beta }}({{x}_{j}}){{\partial }_{\nu }}{{g}_{\alpha \beta }}({{x}_{j}}) \right\rangle
\end{align}
One can alternatively write it in terms of connections
\begin{align}
\left\langle x_{i}^{\mu }\frac{\partial H}{\partial x_{j}^{\nu }} \right\rangle =\frac{1}{\beta }\delta _{\nu }^{\mu }{{\delta }_{ij}}+\frac{1}{\beta }\left\langle x_{i}^{\mu }\Gamma _{\alpha \nu }^{\alpha }({{x}_{j}}) \right\rangle
\end{align}
Note that the momentum part is the same, but with no geometric contribution
\begin{align}
\left\langle p_{i}^{\mu }\frac{\partial H}{\partial p_{j}^{\nu }} \right\rangle =\frac{1}{\beta }\delta _{\nu }^{\mu }{{\delta }_{ij}}
\end{align}
Combining terms, we obtain a new version of the virial theorem
\begin{align}
\left\langle x_{i}^{\mu }\frac{\partial H}{\partial x_{j}^{\nu }} \right\rangle =\left\langle p_{i}^{\mu }\frac{\partial H}{\partial p_{j}^{\nu }} \right\rangle +\frac{1}{\beta }\left\langle x_{i}^{\mu }\Gamma _{\alpha \nu }^{\alpha }({{x}_{j}}) \right\rangle
\end{align}
One can also take a summation over these quantities; however, such a sum will produce a new term if we expand the total energy in terms of the kinetic and potential energies, giving
\begin{align}
  & \sum\limits_{i}{\left\langle x_{i}^{\mu }\frac{\partial H}{\partial x_{i}^{\mu }} \right\rangle }=\sum\limits_{i}{\left\langle x_{i}^{\mu }\frac{\partial V}{\partial x_{i}^{\mu }} \right\rangle }+\sum\limits_{i}{\left\langle x_{i}^{\mu }\frac{\partial E_K}{\partial x_{i}^{\mu }} \right\rangle } \nonumber\\
 & =\sum\limits_{i}{\left\langle x_{i}^{\mu }\frac{\partial V}{\partial x_{i}^{\mu }} \right\rangle }+\frac{1}{2}\sum\limits_{i}{\left\langle {{m}_{i}}{{\partial }_{\mu }}{{g}_{\alpha \beta }}({{x}_{i}})x_{i}^{\mu }\dot{x}_{i}^{\alpha }\dot{x}_{i}^{\beta } \right\rangle }\nonumber\\
&=\sum\limits_{i}{\left\langle x_{i}^{\mu }\frac{\partial V}{\partial x_{i}^{\mu }} \right\rangle }+\sum\limits_{i}{\left\langle {{m}_{i}}{{\Gamma }_{\alpha \beta \mu }}({{x}_{i}})x_{i}^{\mu }\dot{x}_{i}^{\alpha }\dot{x}_{i}^{\beta } \right\rangle }
\end{align}
So the modified virial theorem is
\begin{align}
\sum\limits_{i}{\left\langle x_{i}^{\mu }\frac{\partial V}{\partial x_{i}^{\mu }} \right\rangle }+\sum\limits_{i}{\left\langle {{m}_{i}}{{\Gamma }_{\alpha \beta \mu }}({{x}_{i}})x_{i}^{\mu }\dot{x}_{i}^{\alpha }\dot{x}_{i}^{\beta } \right\rangle }=2\left\langle E_K \right\rangle +\frac{1}{\beta }\sum\limits_{i}{\left\langle x_{i}^{\mu }\Gamma _{\alpha \mu }^{\alpha }({{x}_{i}}) \right\rangle }
\end{align}
\subsubsection{Mechanical version}
One can alternatively study the mechanical version of the virial theorem by taking a time average. One can define the quantity
\begin{align}
G=\sum\limits_{i}{{{g}_{\mu \nu }}({{x}_{i}})p_{i}^{\mu }x_{i}^{\nu }}
\end{align}
So similar with the previous derivation, we take the derivative over $t$ to get
\begin{align}
  & \frac{dG}{dt}=\sum\limits_{i}{{{\partial }_{\lambda }}{{g}_{\mu \nu }}({{x}_{i}}){{m}_{i}}\dot{x}_{i}^{\lambda }\dot{x}_{i}^{\mu }x_{i}^{\nu }}+\sum\limits_{i}{{{g}_{\mu \nu }}({{x}_{i}}){{m}_{i}}\ddot{x}_{i}^{\mu }x_{i}^{\nu }}+\sum\limits_{i}{{{g}_{\mu \nu }}({{x}_{i}}){{m}_{i}}\dot{x}_{i}^{\mu }\dot{x}_{i}^{\nu }} \nonumber\\
 & =\sum\limits_{i}{{{\partial }_{\lambda }}{{g}_{\mu \nu }}({{x}_{i}}){{m}_{i}}\dot{x}_{i}^{\lambda }\dot{x}_{i}^{\mu }x_{i}^{\nu }}+\sum\limits_{i}{{{g}_{\mu \nu }}({{x}_{i}}){{m}_{i}}\ddot{x}_{i}^{\mu }x_{i}^{\nu }}+2E_K
\end{align}
So take the average will give a modified version of the virial theorem,
\begin{align}
\left\langle \sum\limits_{i}{\left( {{\partial }_{\lambda }}{{g}_{\beta \nu }}({{x}_{i}})-{{\Gamma }_{\nu \beta \lambda }}({{x}_{i}}) \right){{m}_{i}}\dot{x}_{i}^{\beta }\dot{x}_{i}^{\lambda }x_{i}^{\nu }}+2E_K \right\rangle =\left\langle \sum\limits_{i}{\frac{\partial V}{\partial x_{i}^{\nu }}x_{i}^{\nu }} \right\rangle
\end{align}
where we use the equation of motion
\begin{align}
{{m}_{i}}({{g}_{\mu \beta }}\ddot{x}_{i}^{\beta }+\dot{x}_{i}^{\beta }\dot{x}_{i}^{\lambda }{{\partial }_{\lambda }}{{g}_{\mu \beta }}({{x}_{i}}))=\frac{{{m}_{i}}}{2}\dot{x}_{i}^{\alpha }\dot{x}_{i}^{\beta }{{\partial }_{\mu }}{{g}_{\alpha \beta }}({{x}_{i}})-\frac{\partial V}{\partial x_{i}^{\mu }}
\end{align}
Using differential geometry identity
\begin{align}
  & {{\Gamma }_{\nu \beta \lambda }}=\frac{1}{2}\left( {{\partial }_{\lambda }}{{g}_{\beta \nu }}+{{\partial }_{\beta }}{{g}_{\lambda \nu }}-{{\partial }_{\nu }}{{g}_{\beta \lambda }} \right) \nonumber\\
 & {{\partial }_{\lambda }}{{g}_{\beta \nu }}-{{\Gamma }_{\nu \beta \lambda }}={{\partial }_{\lambda }}{{g}_{\beta \nu }}-\frac{1}{2}\left( {{\partial }_{\lambda }}{{g}_{\beta \nu }}+{{\partial }_{\beta }}{{g}_{\lambda \nu }}-{{\partial }_{\nu }}{{g}_{\beta \lambda }} \right)=\frac{1}{2}\left( {{\partial }_{\lambda }}{{g}_{\beta \nu }}-{{\partial }_{\beta }}{{g}_{\lambda \nu }}+{{\partial }_{\nu }}{{g}_{\beta \lambda }} \right)={{\Gamma }_{\beta \nu \lambda }}
\end{align}
We get
\begin{align}
\left\langle \sum\limits_{i}{{{\Gamma }_{\beta \nu \lambda }}({{x}_{i}}){{m}_{i}}\dot{x}_{i}^{\beta }\dot{x}_{i}^{\lambda }x_{i}^{\nu }} \right\rangle +2\left\langle E_K \right\rangle =\left\langle \sum\limits_{i}{\frac{\partial V}{\partial x_{i}^{\nu }}x_{i}^{\nu }} \right\rangle
\end{align}
Note here that the mechanical and statistical virial theorems are, in fact, different. The reason is that now in a curved space, the argument of ergodicity is broken. The different points in space are not equally likely to be accessed at late time; the true probability depends sensatively on the shape of the manifold and on the initial positions and momenta of the particles. For our usage, we will use the claim to identify the quantum complexity and statistical entropy, so we could use the statistical virial theorem to derive a relation between two parts of entropy, while some further issues about ergodicity will be commented later. 
\section{A relation between complexities}\label{relation}
\subsection{Quantum/classical correspondense}
After we establish a version of virial theorem in the curved space, we will establish a relation between entropies, and namely, complexities due to the following quantum/classical correspondence established in \cite{Brown:2017jil}.
\\
\\
Consider a $k$-local Hamiltonian with $K$ qubits. The Hamitlonian is defined in the sense of disordered average. The form of it is 
\begin{align}
\hat{H} = \sum\limits_I {{J_I}{\sigma _I}} 
\end{align}
where $I$ runs over all $4^K-1$ Paulis, and $J_I$ is the coupling constant distributed in Gaussian distribution. 
\begin{align}
P(J)\sim \exp(-\frac{1}{2}B_a\sum_I J_I^2)
\end{align}
where $B_a$ defines the variance. Now we are asking what is the computational complexity for an operator $\exp(i\hat{H}t)$. The generic paradigm for complexity evolving with time is given as the following. The computational complexity will firstly increase with time roughly linearly, then it will stay a constant. After a very long time, recurrence will happen and the complexity will decay and grow back. 
\\
\\
\cite{Brown:2017jil} notices a similar behavior should appear for entropies in classical systems. By counting degree of freedom, the dual classical system should have $2^K$ variables. Thus it is necessary to study particle trajectories over the Nielsen's metric construction of quantum computing \cite{Ne0,Ne} on the group manifold $\text{SU}(2^K)$\footnote{It is argued in \cite{Brown:2017jil} that although the dimension of $\text{SU}(2^K)$ is $4^K-1$, because the Hamiltonian has only $2^K$ eigenvalues, the particle is actually moving on a $2^K$ dimensional torus.}. In this metric definition, the metric as a bilinear for Hamiltonian representation of vector fields $\hat{H}_1$ and $\hat{H}_2$ near point $\hat{U}$ is
\begin{align}
\left\langle {{{\hat H}_1},{{\hat H}_2}} \right\rangle  = \frac{{{\rm{Tr}}\left( {{{\hat H}_1}{\cal P}({{\hat H}_2})} \right) + q{\rm{Tr}}\left( {{{\hat H}_1}{\cal Q}({{\hat H}_2})} \right)}}{{{2^K}}}
\end{align}
where $\mathcal{P}$ and $\mathcal{Q}$ are super operators that takes the Hamitonian to the one and two body term components and three or more body components respectively. In this geometry, the complexity is proportional to the geodesic length, or one can also understand it as the action.
\\
\\
For the practical usage, we will consider the Pauli basis. Define
\begin{align}
{\cal G} = {\cal P} + q{\cal Q}
\end{align}
we could define the metric, at a generic point $X$, by
\begin{align}
{g_{\mu \nu }} = \frac{{{\rm{Tr}}(\mu {{\cal G}_X}(\nu ))}}{{{2^K}}}
\end{align}
where $\mu$ and $\nu$ are $K$-qubit Paulis. Here $\mathcal{G}_X$ is defined by
\begin{align}
{{\cal G}_X} = {\cal E}_X^\dag  \circ {\cal G} \circ {{\cal E}_X}
\end{align}
with 
\begin{align}
{{\cal E}_X} = \sum\limits_{j = 0}^{ + \infty } {\frac{{{{( - i{\rm{a}}{{\rm{d}}_X})}^j}}}{{(j + 1)!}}} 
\end{align}
and 
\begin{align}
{\rm{a}}{{\rm{d}}_X}(Y) \equiv \left[ {X,Y} \right]
\end{align}
With this formula, one could study geometric data at arbitrary points, although do an infinite sum is highly non-trivial\footnote{This geometric is non-trivial for even single qubit due to non-commutativity of Paulis although in this case there is no $q$ parameter coming into the metric.}. As a simple application, do fewer expansion in this summation formula, one can obtain some derivatives of the metric to obtain connections, etc.
\\
\\
Imagining a single particle on this group manifold running from the origin, one can study the classical physics of it by given the Hamiltonian $\hat{H}$. One can also notice that the initial velocity components in this classical setup are given by the couplings $J_I$, while the whole initial velocity is given by $K$. Thus, by the claim that complexity is equal to the geodesic length, we claim that initially, the complexity grows as $Kt$. In fact, \cite{Brown:2017jil} conjectures that, the computational complexity at the quantum side should be proportional to the \emph{positional} entropy in such a dual classical system, where the distribution (disorder of $J_I$) naturally defines a classical ensemble. The reason for only \emph{positional} entropy instead of the whole entropy is that the computational complexity is only related to the position from the origin, not the velocity.
\\
\\
Because now we make use of the positional entropy, what is the interpretation for the kinetic entropy? \cite{Brown:2017jil} argues that it should be understood as the Kolmogorov complexity for the Hamiltonian. In fact, this is based on the duality between the velocities and the coupling constants that we have discussed above. Imagine that the coupling $J_I$s are bits $0$ or $1$, so the possibilities of couplings, or namely, the possibilities of bits, should be related to the bit string realization of the Hamiltonian. Thus it is reasonable that the cost to specify the bit string, namely, the Kolmogorov complexity, is connected to the kinetic entropy which is related to the velocities. This argument works for the concept of Kolmogorov complexity for single instance of Hamiltonian, which is not related to the status of the ensemble, and makes it hard to work with. However, here we could generalize the argument to the ensemble averaged version of the Kolmogorov complexity. Considering the probability distribution of the coupling $P(J)$, and the entropy $-\sum P(J)\log P(J)$, under mild assumptions, it is claimed that it is equal to the averaged version of the Kolmogorov complexity $\sum P(J)C_\kappa(J)$, where $C_\kappa(J)$ is the Kolmogorov complexity for single instant.
\\
\\
In the following subsection, we will describe a relationship between Kolmogorov complexity and computational complexity motivated by the virial theorem we describe above. The Kolmogorov complexity is a static object independent of time, but computational complexity could be defined in every instant. However, we will consider the equilibrium case, which correspond to a late time value of computational complexity in which the system is in a thermal bath with inverse temperature $\beta$. 

\subsection{An entropic/complexity relation}
We can decompose the entire entropy in the following way, where we consider a canonical ensemble with inverse temperature $\beta$:
\begin{align}
  & S=-\int{d\Omega \beta H\exp (-\beta H)}={{S}_{K}}+{{S}_{P}} \nonumber\\
 & {{S}_{K}}=-\int{d\Omega \beta E_K\exp (-\beta H)}=-\beta \left\langle E_K \right\rangle   \nonumber\\
 & {{S}_{P}}=-\int{d\Omega \beta V\exp (-\beta H)}=-\beta \left\langle V \right\rangle
\end{align}
To make this relation work with tractible computations, we make the assumption that \emph{all coordinates $x^\mu$ are sufficiently close to the origin $x^\mu=0$}\footnote{This assumption is motivated by the fact that geodesics on the Nielsen complexity geometry are only fully understood in the case where the geodesic distances between points are quite short, to avoid troublesome conjugate point ambiguities in the calculation. It could be, however, that these relations would generalize to larger geodesic lengths if further techniques to understand longer geodesics in the Nielsen complexity geometry are developed.}. In this case, we could write $x$ as $\Delta x$. Then, defining $V(x_i^\mu=0)=0$, then we could have\footnote{In this application, there is no index $i$ specifying particles because there is only one single particle, or namely all degree of freedoms are understood as different coordinates on the group manifold, and the mass is set to one.} 
\begin{align}
\frac{1}{2}\left\langle {\Delta {x^\mu }\Delta {x^\nu }\frac{{\partial V}}{{\partial {x^\mu }\partial {x^\nu }}}} \right\rangle  + \left\langle {\Delta {x^\mu }\frac{{\partial V}}{{\partial {x^\mu }}}} \right\rangle  = \left\langle {V(\Delta x)} \right\rangle 
\end{align}
Where we keep it to the second order. In this limit, we obtain
\begin{align}
\left\langle V \right\rangle  - \frac{1}{2}\left\langle {\Delta {x^\mu }\Delta {x^\nu }\frac{{\partial V}}{{\partial {x^\mu }\partial {x^\nu }}}} \right\rangle  + \left\langle {{\Gamma _{\alpha \beta \mu }}(\Delta x)\Delta {x^\mu }\Delta {{\dot x}^\alpha }\Delta {{\dot x}^\beta }} \right\rangle  = 2\left\langle E_K \right\rangle  + \left\langle {\frac{1}{\beta }\Delta {x^\mu }\Gamma _{\alpha \mu }^\alpha (\Delta x)} \right\rangle 
\end{align}
So we have
\begin{align}
&\frac{{2{S_K}}}{\beta } - \frac{{{S_P}}}{\beta } = \frac{1}{\beta }\left\langle {\Delta {x^\mu }\Gamma _{\alpha \mu }^\alpha (\Delta x)} \right\rangle  - \left\langle {{\Gamma _{\alpha \beta \mu }}(\Delta x)\Delta {x^\mu }\Delta {{\dot x}^\alpha }\Delta {{\dot x}^\beta }} \right\rangle  + \frac{1}{2}\left\langle {\Delta {x^\mu }\Delta {x^\nu }{\partial _{\mu \nu }}V} \right\rangle \nonumber\\
&= \frac{1}{\beta }\Gamma _{\alpha \mu }^\alpha \left\langle {\Delta {x^\mu }} \right\rangle  - {\Gamma _{\alpha \beta \mu }}\left\langle {\Delta {x^\mu }\Delta {{\dot x}^\alpha }\Delta {{\dot x}^\beta }} \right\rangle \nonumber\\
&+ \frac{1}{2}\left\langle {\Delta {x^\mu }\Delta {x^\nu }{\partial _{\mu \nu }}V} \right\rangle  + \frac{1}{\beta }{\partial _\nu }\Gamma _{\alpha \mu }^\alpha \left\langle {\Delta {x^\mu }\Delta {x^\nu }} \right\rangle  - {\partial _\nu }{\Gamma _{\alpha \beta \mu }}\left\langle {\Delta {x^\mu }\Delta {x^\nu }\Delta {{\dot x}^\alpha }\Delta {{\dot x}^\beta }} \right\rangle 
\end{align}
where the last formula is expanded around $x=0$\footnote{Here we write, for instance, $\Gamma_{\alpha\tau}^\alpha \equiv \Gamma_{\alpha\tau}^\alpha(0) $ means quantities evaluated at the origin for short.}, and we expand the result at the order $\mathcal{O}(\Delta x)$. This relationship could be simplified further in the Nielsen's geometry \cite{Ne}, which could show that the linear order has zero contribution. We have
\begin{align}
{{\Gamma }_{\mu \sigma \tau }}=\frac{i}{{{2}^{K+1}}}\text{Tr}\left( \mu \left( \left[ \sigma ,\mathcal{G}(\tau ) \right]+\left[ \tau ,\mathcal{G}(\sigma ) \right] \right) \right)
\end{align}
where the indices $\sigma,\tau$ etc denote the possibilities of Paulis. One can also derive the connection with one upper index
\begin{align}
\Gamma _{\sigma \tau }^{\rho }=\frac{i}{{{2}^{K+1}}}\text{Tr}\left( \mathcal{F}(\rho )\left( \left[ \sigma ,\mathcal{G}(\tau ) \right]+\left[ \tau ,G(\sigma ) \right] \right) \right)
\end{align}
where $\mathcal{F}=\mathcal{G}^{-1}$. By cyclic property of trace, we get
\begin{align}
&\Gamma _{\alpha \mu }^\alpha=0\nonumber\\
&{\Gamma _{\alpha \beta \mu }}=-{\Gamma _{\beta \alpha \mu }}
\end{align}
So we obtain a vanishing leading order result. Thus, we have to look at next leading order
\begin{align}
\frac{{2{S_K}}}{\beta } - \frac{{{S_P}}}{\beta }= \frac{1}{2}\left\langle {\Delta {x^\mu }\Delta {x^\nu }{\partial _{\mu \nu }}V} \right\rangle  + \frac{1}{\beta }{\partial _\nu }\Gamma _{\alpha \mu }^\alpha \left\langle {\Delta {x^\mu }\Delta {x^\nu }} \right\rangle  - {\partial _\nu }{\Gamma _{\alpha \beta \mu }}\left\langle {\Delta {x^\mu }\Delta {x^\nu }\Delta {{\dot x}^\alpha }\Delta {{\dot x}^\beta }} \right\rangle 
\end{align}
Now let us simplify it further after the following assumptions. We assume 
\begin{align}
&\left\langle {\Delta {x^\mu }\Delta {x^\nu }} \right\rangle  = {\delta ^{\mu \nu }}\Delta {L^2}\nonumber\\
&\left\langle {\Delta {{\dot x}^\mu }\Delta {{\dot x}^\nu }} \right\rangle  = {\delta ^{\mu \nu }}{v^2}\nonumber\\
&\left\langle {\Delta {x^\mu }\Delta {x^\nu }\Delta {{\dot x}^\alpha }\Delta {{\dot x}^\beta }} \right\rangle  = \left\langle {\Delta {x^\mu }\Delta {x^\nu }} \right\rangle \left\langle {\Delta {{\dot x}^\alpha }\Delta {{\dot x}^\beta }} \right\rangle 
\end{align}
These statistical assumptions are based on both fluctuations of velocities and coordinates are Gaussian, and there is no correlation between position and momentum\footnote{The similar Gaussianity assumption follows from the distribution of the coupling is also made in \cite{Brown:2017jil}. It will be interesting to extend our research for non-Gaussian case in the future, and it might be also related to the complexity for Gaussian states \cite{Jefferson:2017sdb,Hackl:2018ptj}.}. Based on this, the formula could be simplified as 
\begin{align}
\frac{{2{S_K}}}{\beta } - \frac{{{S_P}}}{\beta } = \frac{1}{2}\left( {\sum\limits_\mu  {{\partial _{\mu \mu }}V} } \right)\Delta {L^2} + \frac{1}{\beta }\left( {\sum\limits_\mu  {{\partial _\mu }\Gamma _{\alpha \mu }^\alpha } } \right)\Delta {L^2} - \left( {\sum\limits_{\mu \alpha } {{\partial _\mu }{\Gamma _{\alpha \alpha \mu }}} } \right)\Delta {L^2}{v^2}
\end{align}
Thus, based on the conjectures  in \cite{Brown:2017jil}, the ensemble average of computational (Kolmogorov) complexity of a disordered $k$-local quantum system, should be proportional to positional (kinetic) part of statistical entropy of a dual statistical gas living in the group manifold. From the entropy relation derived above, we arrive at a direct relation between computational and Kolmogorov complexity. The relation is pedagogically
\begin{align}
\text{Kolmogorov complexity}=\text{Computational complexity}+\text{Corrections}
\end{align}
where the corrections can be computed directly from Nielsen's geometry and the potential.
\subsection{Analysis}
We will consider the large $K$ limit, and some small qubits examples to make some claims on this relationship. Before the precise investigations, we could give some generic comments.
\begin{itemize}
\item For extremely small $\Delta L$, there is nearly no correction between two entropies, where we could claim that they are nearly proportional, at least at the leading order of $\Delta L$. However, one can expect that when the number of qubits are large, it is very easy to achieve a large number in the RHS, measuring the difference of two entropies (complexities), because the number of degree of freedom increases exponentially. This is consistent with related arguments in \cite{Brown:2017jil} about exponential dominance of the computational complexity. Moreover, assuming a large dominance for computational complexity than Kolmogorov complexity, we obtain a bound
\begin{align}
\frac{\beta }{2}\left( {\sum\limits_\mu  {{\partial _{\mu \mu }}V} } \right)\Delta {L^2} + \left( {\sum\limits_\mu  {{\partial _\mu }\Gamma _{\alpha \mu }^\alpha } } \right)\Delta {L^2} - \left( {\sum\limits_{\mu \alpha } {{\partial _\mu }{\Gamma _{\alpha \alpha \mu }}} } \right)\Delta {L^2}\beta {v^2}  \gtrsim 1
\end{align}
\item Secondly, if we treat the distance variation $\Delta L$ to be small enough, the difference between mechanical and statistical version of the virial theorem is tiny. The disappearing of the different term is a recovery of ergodicity in the statistical ensemble we consider. In this sense, the ensemble average is equal to the long term average at the leading order. 
\item This formula only works for equilibrium, where we have a thermal bath with temperature $\beta$, or some ergodic states with an effective temperature $\beta$ if we treat $\Delta L$ to be small enough. Thus, this formula cannot show the time dependence of the computational complexity, although it should work for some ergodic states where the instant ensemble average is equal to the long term average in the small $\Delta L$ limit.
\item This formula is consistent when sending $\beta\to\infty$, where trajectories move slowly so both side will be suppressed.
\end{itemize}

\subsubsection{Sufficiently large $K$}
The most interesting case might be the large $K$ limit, where we could make some generic analysis based on the geometry of the group manifold. Firstly, consider two double summation terms
\begin{align}
- \left( {\sum\limits_{\mu \alpha } {{\partial _\mu }{\Gamma _{\alpha \alpha \mu }}} } \right)\Delta {L^2}{v^2}~,~~~~~\frac{1}{\beta }\left( {\sum\limits_\mu  {{\partial _\mu }\Gamma _{\alpha \mu }^\alpha } } \right)\Delta {L^2}
\end{align}
Because there are $4^K-1$ Paulis in total, so we expect that it will scale as $\mathcal{O}(16^K)$. We could make a more detailed estimation here. We note that 
\begin{align}
{\partial _\mu }{\Gamma _{\alpha \alpha \mu }} = \frac{1}{2}{\partial _{\mu \mu }}{g_{\alpha \alpha }}
\end{align}
where the derivative of the metric could be given in the Pauli basis formula from the metric. Expanding the expression we have
\begin{align}
\frac{1}{2}{\partial _{\mu \mu }}{g_{\alpha \alpha }} =  - \frac{2}{{{2^K}}}{\rm{Tr}}\left( {\frac{{{\rm{a}}{{\rm{d}}_\mu }{\rm{a}}{{\rm{d}}_\mu }\alpha }}{6}{\cal G}(\alpha )} \right) - \frac{1}{{{2^K}}}{\rm{Tr}}\left( {\frac{{{\rm{a}}{{\rm{d}}_\mu }\alpha }}{4}{\cal G} \circ {\rm{a}}{{\rm{d}}_\mu }(\alpha )} \right)
\end{align}
One could define the inner product
\begin{align}
\left\langle {A,B} \right\rangle  = \frac{1}{{{2^K}}}{\rm{Tr}}\left( {A\mathcal{G}(B)} \right)
\end{align}
then the formula has been re-expressed as 
\begin{align}
{\partial _{\mu \mu }}{g_{\alpha \alpha }} =  - \frac{2}{3}\left\langle {\left[ {\mu ,\left[ {\mu ,\alpha } \right]} \right],\alpha } \right\rangle  - \frac{1}{2}\left\langle {\left[ {\mu ,\alpha } \right],\left[ {\mu ,\alpha } \right]} \right\rangle 
\end{align}
The second term will count for weights of $[\mu,\alpha]$ for non-commuting pairs. This term will scale as $16^K$. Similar thing happens for the first term. Thus in general we estimate
\begin{align}
- \left( {\sum\limits_{\mu \alpha } {{\partial _\mu }{\Gamma _{\alpha \alpha \mu }}} } \right)\Delta {L^2}{v^2}\sim 16^K \Delta {L^2}{v^2} \times \mathcal{O}(1)
\end{align}
Then we move to another double sum geometric term
\begin{align}
\frac{1}{\beta }\left( {\sum\limits_\mu  {{\partial _\mu }\Gamma _{\alpha \mu }^\alpha } } \right)\Delta {L^2}
\end{align}
By definition
\begin{align}
{\partial _\mu }\Gamma _{\alpha \mu }^\alpha  = \frac{1}{2}{g^{\alpha \nu }}{\partial _{\mu \mu }}{g_{\alpha \nu }} + \frac{1}{2}({\partial _\mu }{g^{\alpha \nu }})({\partial _\mu }{g_{\alpha \nu }})
\end{align}
The first term is nothing but dividing an $\mathcal{O}(1)$ constant, $1$ or $1/q$ in each valid term of the summation. The second term gives
\begin{align}
({\partial _\mu }{g_{\alpha \nu }})({\partial _\mu }{g^{\alpha \nu }}) = \frac{1}{4}\frac{{{{({q_\alpha } - {q_\nu })}^2}}}{{{q_\alpha }{q_\nu }}}{\left( {\frac{{{\rm{Tr}}\left( {\left[ {\alpha ,\nu } \right]\mu } \right)}}{{{2^K}}}} \right)^2}
\end{align}
Where $q_\alpha$ means $1$ for one, and two body Paulis, and $q$ for three body and more. Thus we have
\begin{align}
\sum\limits_\mu  {({\partial _\mu }{g_{\alpha \nu }})({\partial _\mu }{g^{\alpha \nu }})}  \sim {16^K} \times \frac{{{{(q - 1)}^2}}}{q} \times \mathcal{O}(1)
\end{align}
So we conclude that 
\begin{align}
\frac{1}{\beta }\left( {\sum\limits_\mu  {{\partial _\mu }\Gamma _{\alpha \mu }^\alpha } } \right)\Delta {L^2} \sim {16^K} \times \frac{{\Delta {L^2}}}{\beta } \times \mathcal{O}(1)
\end{align}
for generic $q$. Finally, we comment on the potential term. We find that both geometric terms are double sum, while the potential term is single sum. Thus it is at most proportional to $4^K$,
\begin{align}
\frac{1}{2}\left( {\sum\limits_\mu  {{\partial _{\mu \mu }}V} } \right)\Delta {L^2} \le {4^K} \times {\max _\mu }\left( {{\partial _{\mu \mu }}V} \right) \times \Delta {L^2}
\end{align}
Thus in the large $K$ limit, we could drop out the potential term, thus the relationship should look like 
\begin{align}
\frac{{2{S_K}}}{\beta } - \frac{{{S_P}}}{\beta } \sim {16^K} \times \frac{{\Delta {L^2}}}{\beta } \times \mathcal{O}(1) + {16^K} \times \Delta {L^2}{v^2} \times \mathcal{O}(1)
\end{align}
Now we make further estimations over the parameter $\beta$ and $v$ for large $K$. We have \cite{Brown:2017jil}
\begin{align}
&\beta  \sim {K^{k - 1}}\nonumber\\
&{v^2} \sim {J^2} \sim \frac{{k!}}{{{3^k}{K^{k - 1}}}}
\end{align}
for $k$-local system, and the later is given by the velocity-coupling duality \cite{Brown:2017jil}. So the relationship is simplified further by
\begin{align}
2{S_K} - {S_P}\sim{16^K}\Delta {L^2} \times {\cal O}(1)
\end{align}
Finally, we make a comment that the leading dependence might also be find-tuned by solving specific $q$ for given $K$. In these cases, the dependence over $K$ for those correction terms is weaker, or moreover, these $\Delta L^2$ dependence could even be cancelled. It will be interesting to study how it could happen in general, and the relationship between fine-tuning and physics problems, like ergodicity.
\subsubsection{Fewer qubit examples}
We will list $K=1,2,3$ qubits here as examples. 
\begin{itemize}
\item For $K=1$ we have
\begin{align}
\sum\limits_{\mu \alpha } {{\partial _\mu }{\Gamma _{\alpha \alpha \mu }}}  = \sum\limits_\mu  {{\partial _\mu }\Gamma _{\alpha \mu }^\alpha }  =  - 2
\end{align}
Thus the relationship looks like 
\begin{align}
2{S_K} - {S_P} = \frac{\beta }{2}\left( {\sum\limits_\mu  {{\partial _{\mu \mu }}V} } \right)\Delta {L^2} + 2({v^2}\beta  - 1)\Delta {L^2}
\end{align}
\item For $K=2$ we have
\begin{align}
\sum\limits_{\mu \alpha } {{\partial _\mu }{\Gamma _{\alpha \alpha \mu }}}  = \sum\limits_\mu  {{\partial _\mu }\Gamma _{\alpha \mu }^\alpha }  =  - 40
\end{align}
Thus the relationship looks like
\begin{align}
2{S_K} - {S_P} = \frac{\beta }{2}\left( {\sum\limits_\mu  {{\partial _{\mu \mu }}V} } \right)\Delta {L^2} + 40({v^2}\beta  - 1)\Delta {L^2}
\end{align}
\item For $K=3$ we have
\begin{align}
&\sum\limits_{\mu \alpha } {{\partial _\mu }{\Gamma _{\alpha \alpha \mu }}}  =-384 - 288 q\nonumber\\
&\sum\limits_{\mu \alpha } {{\partial _\mu }\Gamma _{\alpha \mu }^\alpha } =-672
\end{align}
Notice that in the second term, the $q$ dependence has been cancelled. Thus the relationship looks like
\begin{align}
2{S_K} - {S_P} = \frac{\beta }{2}\left( {\sum\limits_\mu  {{\partial _{\mu \mu }}V} } \right)\Delta {L^2} + ((384 + 288q){v^2}\beta  - 672)\Delta {L^2}
\end{align}

\end{itemize}

\section{Conclusion and discussion}\label{con}
In this paper, we study a possible consequence of conjectures in \cite{Brown:2017jil} for identifying complexities in the $k$-local disordered Hamiltonian and classical systems of particulate particles living in the group manifold of unimodular matrices. After a discussion of the virial theorem in a general curved space, we arrive at a nontrivial relation between Kolmogorov complexity and computational complexity by identifying complexities with entropies.
\\
\\
Finally, we will discuss possibilities for future research.
\begin{itemize}
\item It is reasonable to try connecting this work to existing works about holographic complexity and rigorous definition of complexities in quantum field theory, for instance, discuss the classical correspondence for unitaries that could prepare Gaussian states \cite{Jefferson:2017sdb,Hackl:2018ptj}, and what is the implications for virial theorems there.
\item One could also generalize this work to higher orders, namely, the larger deviation $\mathcal{O}(\Delta L^3)$ to ask what is the geometric corrections, and address some physically interesting questions, like ergodicity.
\item One can try simulating Nielsen geometry by solving numerical differential equations, in the classical and future quantum computers.
\item It would be interesting to study such an argument from the complexity theoretic and quantum resource theoretic points of view (where, the Kolmogorov complexity-computational complexity decomposition, or moreover, the complexity-uncomplexity decomposition, claimed in \cite{Brown:2017jil}, may have a meaningful interpretation as quantum resource).
\item It would also be interesting to work out some specific chaotic examples for $k$-local disordered Hamiltonians (like the SYK model \cite{Sachdev:1992fk,A1,A2,A3,A4,Maldacena:2016hyu}\footnote{There are some related discussions in previous work by Roberts and Yoshida \cite{Roberts:2016hpo}, which precisely relates the quantum 2-R\'{e}nyi entropy and the quantum complexity by \emph{frame potential}, which is a quantum information theory quantity measuring the average of out-of-time-ordered correlators. See also some related discussions in \cite{Cotler:2017jue,Hunter-Jones:2017crg} for applications in random matrix theory and SYK-like models.}) to verify validity of the statement in practice, as perhaps a nontrivial check to the conjectures of \cite{Brown:2017jil}.

\end{itemize}

\section*{Acknowledgments}
We thank Elizabeth Crosson, Beni Yoshida, Nicole Yunger Halpern and Shangnan Zhou for helpful discussions. We thank the anonymous JHEP referee for valuable communications. NB is supported by the National Science Foundation, under grant number 82248-13067-44-PHPXH. JL is supported in part by the Institute for Quantum Information and Matter (IQIM), an NSF Physics Frontiers Center (NSF Grant PHY-1125565) with support from the Gordon and Betty Moore Foundation (GBMF-2644), and by the Walter Burke Institute for Theoretical Physics.

\end{document}